\pdfoutput=1
\RequirePackage{ifpdf}
\ifpdf 
\documentclass[pdftex]{sigma}
\else
\documentclass{sigma}
\fi

\DeclareMathOperator{\Ad}{Ad}
\DeclareMathOperator{\ad}{ad}
\DeclareMathOperator{\id}{id}
\DeclareMathOperator{\tr}{tr}
\DeclareMathOperator{\diag}{diag}
\newcommand{\abs}[1]{|#1|}
\numberwithin{equation}{section}

\begin{document}

\allowdisplaybreaks

\renewcommand{\thefootnote}{$\star$}

\renewcommand{\PaperNumber}{053}

\FirstPageHeading

\ShortArticleName{Towards Non-Commutative Deformations of Relativistic Wave Equations}

\ArticleName{Towards Non-Commutative Deformations\\
of Relativistic Wave Equations in 2+1 Dimensions\footnote{This paper is a~contribution to the Special Issue on
Deformations of Space-Time and its Symmetries.
The full collection is available at \href{http://www.emis.de/journals/SIGMA/space-time.html}
{http://www.emis.de/journals/SIGMA/space-time.html}}}

\Author{Bernd J.~SCHROERS~$^\dag$ and Matthias WILHELM~$^\ddag$}

\AuthorNameForHeading{B.J.~Schroers and M.~Wilhelm}

\Address{$^\dag$~Department of Mathematics and Maxwell Institute for Mathematical Sciences,\\
\hphantom{$^\dag$}~Heriot-Watt University, Edinburgh EH14 4AS, UK}
\EmailD{\href{mailto:b.j.schroers@hw.ac.uk}{b.j.schroers@hw.ac.uk}}
\URLaddressD{\url{http://www.macs.hw.ac.uk/~bernd/}}

\Address{$^\ddag$~Institut f\"ur Mathematik und Institut f\"ur Physik, Humboldt-Universit\"at zu Berlin,\\
\hphantom{$^\ddag$}~IRIS-Adlershof, Zum Gro\ss{}en Windkanal 6, 12489 Berlin, Germany}
\EmailD{\href{mailto:mwilhelm@physik.hu-berlin.de}{mwilhelm@physik.hu-berlin.de}}

\ArticleDates{Received February 28, 2014, in f\/inal form May 09, 2014; Published online May 20, 2014}

\Abstract{We consider the deformation of the Poincar\'e group in 2+1 dimensions into the quantum double of the Lorentz
group and construct Lorentz-covariant momentum-space formulations of the irreducible representations describing massive
particles with spin~0, $\frac12$ and~1 in the deformed theory.
We discuss ways of obtaining non-commutative versions of relativistic wave equations like the Klein--Gordon, Dirac and
Proca equations in 2+1 dimensions by applying a~suitably def\/ined Fourier transform, and point out the relation between
non-commutative Dirac equations and the exponentiated Dirac operator considered by Atiyah and Moore.}

\Keywords{relativistic wave equations; quantum groups; curved momentum space; non-commutative spacetime}

\Classification{83A99; 81R20; 81R50; 81R60}

\renewcommand{\thefootnote}{\arabic{footnote}}
\setcounter{footnote}{0}

\section{Introduction}

It is well-known that the important linear wave equations of relativistic physics can be obtained by
Fourier transforming the irreducible representations of the Poincar\'e group.
The Klein--Gordon, Dirac and Proca equations, for example, are Fourier transforms of momentum-space constraints for,
respectively, spin 0, $\frac 1 2 $ and 1 in Wigner's classif\/ication of irreducible Poincar\'e representations in terms
of mass and spin~\cite{BR, Sternberg}.

In this paper, we discuss this picture for the case of (2+1)-dimensional Minkowski space, and then consider
a~deformation of it where the Poincar\'e symmetry is deformed into a~non-cocommutative quantum group, namely the quantum
double of the Lorentz group in 2+1 dimensions, or Lorentz double for short~\cite{BM,BMS,KM}.
The deformation involves a~parameter of dimension inverse mass, and deforms f\/lat momentum space of ordinary special
relativity into anti-de Sitter space; in 2+1 dimensions, this happens to be isometric to the identity component of the
Lorentz group.

\looseness=-1
The Lorentz double plays an important role in (2+1)-dimensional quantum gravity~\cite{BM,BMS,MS1,MS2,Schroers}.
In that context, the deformation parameter is related to Newton's constant.
We will not discuss the gravitational interpretation much in this paper and refer to the review~\cite{SchroersCracow}
for details and references.
Instead we focus on general, structural features of our (2+1)-dimensional example, treating it as a~case study of more
general deformations of momentum spaces to curved manifolds.
\looseness=-3

Such deformations have been considered in various guises and with dif\/ferent motivations in the physics literature.
Early considerations of curvature in momentum space include the work of Born~\cite{Born} on a~duality between position
and momentum, and also the inf\/luential paper by Snyder~\cite{Snyder} where momentum space is taken to be the de Sitter
manifold.
Majid's bicrossproduct construction~\cite{MajidPhD,Majidbook} provides a~mathematical framework for deforming spacetime
symmetries which naturally accommodates curved momentum space.
The famous deformation of Poincar\'e symmetry into the $\kappa$-Poincar\'e algebra~\cite{kappa} was later seen to f\/it
into this framework~\cite{MaRu}.
In recent years, phenomenological implications of these ideas have been explored extensively under the headings of
`doubly special relativity'~\cite{DSR} and `relative locality'~\cite{relloc}.

In our (2+1)-dimensional theory, position coordinates, which are translation generators in momentum space, no longer
commute.
Instead, they satisfy the Lie algebra of the Lorentz group in 2+1 dimensions and act on the Lorentz group-valued momenta
by inf\/initesimal multiplication (see~\cite{MW} for an early discussion of this point and~\cite{SchroersCracow} for
a~review and further references).
One therefore expects that Fourier transforming the irreducible representations of the Lorentz double, where states are
functions on momentum spaces, will lead to covariant wave equations on a~non-commutative spacetime.

In this paper, we take the f\/irst steps towards realising that expectation.
Our treatment follows a~similar discussion of the Euclidean situation in~\cite{MS}, which is our main reference.
As we shall see, the Lorentzian situation is considerably more involved than the Euclidean case.

We begin, in Section~\ref{sec: relativistic wave equations},
by writing the unitary irreducible representations (UIR's) of the usual (2+1)-dimensional
Poinar\'e group in a~covariant form that allows us to obtain relativistic wave equations via Fourier transform.
Even though the wave equations we obtain are the standard Klein--Gordon, Dirac and Proca equation in 2+1 dimensions, our
method for obtaining them does not appear to have been considered in the literature.

A full classif\/ication the UIR's of the Poincar\'e group in 2+1 dimensions was f\/irst given by Binegar in~\cite{Binegar},
where he also discusses the possibility~-- and dif\/f\/iculties~-- of writing the UIR's in terms of f\/ields on Minkowski
space obeying covariant wave equations.
A complete analysis of relativistic wave equations in 2+1 dimensions is given in~\cite{Gitman} from the point of view of
generalised regular representations.
Our approach gives a~less general treatment of the Poincar\'e UIR's, but maintains the link via Fourier transform
between momentum space and position space.
This link is essential in our derivation of non-commutative wave equations from irreducible representations of the
Lorentz double in subsequent sections.

In Section~\ref{sec: deforming}, we review the representation theory of the Lorentz double and then adapt the covariantisation procedure
developed in Section~\ref{sec: relativistic wave equations}
to the irreducible representations of the Lorentz double, still following the treatment of the
Euclidean situation in~\cite{MS}.
Section~\ref{noncommutative Fourier transformation} is concerned with the translation of the momentum space constraints
into a~spacetime picture.
We use two dif\/ferent kinds of Fourier transform to obtain wave equations from the covariant momentum constraints.
One is a~Fourier transform adapted to quantum groups~\cite{KeMa, Majidbook} where `plane waves' are elements of the
Lorentz group and the resulting wave equations are def\/ined on the (suitably completed) universal enveloping algebra of
the Lie algebra of the Lorentz group.
The second is a~group Fourier transform which, in the model considered here, leads to wave equations for functions on
$\mathbb{R}^3$ with a~certain $\star$-product~\cite{FL1,FL2,FM,GOR,IS,SS2,SS}.
We discuss the relationship between the various notions of Fourier transform and point out an interesting connection
with the exponentiated Dirac operator recently proposed by Atiyah and Moore in~\cite{AM}.

As a~caveat we should say that our treatment in Section~\ref{noncommutative Fourier transformation}
is far from complete; it is designed to point out interesting
questions posed by the results of our Section~\ref{sec: deforming} and to prepare the ground for tackling them.
Examples of such questions are discussed in our f\/inal Section~\ref{Section5}, which contains our conclusion and outlook.

\section{Relativistic wave equations in 2+1 dimensions}\label{sec: relativistic wave equations}

\subsection{Conventions and notation}\label{Poincare group}

We denote (2+1)-dimensional Minkowski space by $\mathbb{R}^{2,1}$ and use the `mostly minus' convention for the
Minkowski metric $\eta= \diag(1,-1,-1)$.
We write elements of $\mathbb{R}^{2,1}$ as $x,y,\ldots$ with $x=(x^0,x^1,x^2)$ and
\begin{gather*}
\eta(x,y)=\eta_{ab} x^a y^b=x^0y^0-x^1y^1-x^2y^2.
\end{gather*}
Latin indices range over $0$, $1$, $2$ and summation over repeated indices is implied.

The group of linear transformations of $\mathbb{R}^{2,1}$ that leave $\eta$ invariant is the Lorentz group $L_3={\rm O}(2,1)$.
It has four connected components.
We are mainly interested in the identity component~-- the subgroup of proper orthochronous Lorentz transformations,
denoted $L_3^{+\uparrow}$.

The group of af\/f\/ine transformations that leave the Minkowski distance $\eta(x-y,x-y)$ invariant is the semidirect
product $L_3 \ltimes \mathbb{R}^{3}$ of the Lorentz group with the abelian group of translations.
We call it the extended Poincar\'{e} group.
Its identity component is the Poincar\'{e} group, which we denote as
\begin{gather*}
P_3 = L_3^{+\uparrow} \ltimes \mathbb{R}^{3}.
\end{gather*}
For the semidirect product we use the conventions of~\cite{MS}, which allow for an easy extension to the quantum group
deformation in the next section but are dif\/ferent from those mostly used in the physics literature.
In our conventions, the product of $(\Lambda_1, a_1), (\Lambda_2, a_2) \in P_3$ is given by
\begin{gather*}
(\Lambda_1, a_1)(\Lambda_2, a_2)=(\Lambda_1\Lambda_2, \Lambda_2 a_1 + a_2).
\end{gather*}
One advantage of this convention is that the ordering of the elements can be interpreted as a~factorisation:
$(\Lambda,a)=(\Lambda, 0)(\text{I}, a)$, where $\text{I}$ is the identity in ${\rm O}(2,1)$.

The action of $(\Lambda,a) \in P_{3}$ on the Minkowski space is then the right action
\begin{gather*}
(\Lambda,a): \ x \mapsto x \triangleleft(\Lambda,a)= \Lambda x + a.
\end{gather*}

For a~full classif\/ication of possible excitations in (2+1)-dimensional relativistic physics, including the anyonic ones,
one needs to study the projective UIR's of $P_3$.
These are given by the ordinary UIR's of the universal covering group of $P_3$, which are studied in detail
in~\cite{Grigore}.
Wave equations for anyonic wave functions with inf\/initely many components are investigated in~\cite{JN}.
In this paper, we work with the double cover ${\rm SL}(2,\mathbb{R})$ of $L_3^{+\uparrow}$ and hence the double cover
$\tilde{P}_{3}={\rm SL}(2,\mathbb{R}) \ltimes \mathbb{R}^{3}$ of the Poincar\'{e} group.
The main reason for this is the convenience of working with $2
\times
2$ matrices, and an easier link with the existing literature on the Lorentz double, which mostly uses a~formulation
based on ${\rm SL}(2,\mathbb{R})$.
Note also that, in 3+1 dimensions, the double cover of the Poincar\'e group is the universal cover.

It turns out to be natural and convenient to interpret the translation group $\mathbb{R}^{3}$ as the vector space
$\mathfrak{sl}(2,\mathbb{R})^*$ dual to $\mathfrak{sl}(2,\mathbb{R})$.
Then $\tilde{P}_{3}={\rm SL}(2,\mathbb{R}) \ltimes \mathfrak{sl}(2,\mathbb{R})^*$, where ${\rm SL}(2,\mathbb{R})$ acts on
$\mathfrak{sl}(2,\mathbb{R})^*$ via the coadjoint action.
The right-action of $(g,a)\in \tilde{P}_{3}$ on Minkowski space $\mathfrak{sl}(2,\mathbb{R})^*$ is then given by
\begin{gather*}
(g,a): \  \mathfrak{sl}(2,\mathbb{R})^* \ni x \mapsto x \triangleleft(g,a)= \Ad_g^* x + a.
\end{gather*}
This action preserves the Minkowski metric $\eta$ on $\mathfrak{sl}(2,\mathbb{R})^*$.\footnote{We can think of $\eta$ as
being induced by the Killing form on the dual $(\mathfrak{sl}(2,\mathbb{R})^*)^*$, but this is not essential in the
following.
}

The Lie algebra $\mathfrak{p}_{3} =\mathfrak{sl}(2,\mathbb{R}) \ltimes \mathfrak{sl}(2,\mathbb{R})^*$ is six
dimensional, with translation genera\-tors~$P_0$,~$P_1$ and $P_2$, rotation generator $J_0$ and boost generators $J_1$ and
$J_2$.
They satisfy the commutation relations:
\begin{gather}\label{def commutation relations Poincare}
[J_a,J_b]= \epsilon_{abc}J^c,
\qquad
[J_a,P_b]= \epsilon_{abc}P^c,
\qquad
[P_a,P_b]= 0,
\qquad
a,b=0,1,2,
\end{gather}
where indices are raised via the inverse Minkowski metric $\eta^{ab}$ and $\epsilon_{abc}$ is the totally antisymmetric
tensor in three dimensions normalised such that $\epsilon_{012}=1$.
We are using conventions where the structure constants in the Lie algebra are real.
This has the advantage that we can exponentiate to obtain group elements without needing to insert the imaginary
unit~$i$.
Our conventions dif\/fer from those in~\cite{MS} in this respect.

The vector spaces $ \mathfrak{sl}(2,\mathbb{R}) $ and $\mathfrak{sl}(2,\mathbb{R})^*$, which make up $\mathfrak{p}_{3}$,
are in duality, and the natural pairing between them is invariant and non-degenerate.
This pairing plays an important role in the Chern--Simons formulation of 2+1 gravity~\cite{AT, SchroersCracow,Witten},
where it is normalised via Newton's constant~$G$:
\begin{gather}\label{def bilinear form}
\langle J_a,P_b \rangle = \frac{1}{8\pi G} \eta_{ab},
\qquad
\langle J_a,J_b\rangle=\langle P_a,P_b\rangle = 0.
\end{gather}

\subsection[Irreducible unitary representations of $\tilde{P}_{3}$]
{Irreducible unitary representations of $\boldsymbol{\tilde{P}_{3}}$}

The UIR's of $\tilde{P}_{3}$ are classif\/ied in terms of ${\rm SL}(2,\mathbb{R})$ orbits in $(\mathfrak{sl}(2,\mathbb{R})^*)^*$
together with UIR's of associated stabiliser groups~\cite{BR}.
Since $(\mathfrak{sl}(2,\mathbb{R})^*)^* = \mathfrak{sl}(2,\mathbb{R})$, these orbits are nothing but adjoint orbits of
${\rm SL}(2,\mathbb{R})$.
The following is a~convenient basis of $ \mathfrak{sl}(2,\mathbb{R})$, whose detailed properties are summarised in
Appendix~\ref{conventions}:
\begin{gather}\label{def t_a's}
t^0=\frac{1}{2}
\begin{pmatrix}
\phantom{-} 0 & 1
\\
-1 & 0
\end{pmatrix},
\qquad
t^1=\frac{1}{2}
\begin{pmatrix}
1 & \phantom{-}0
\\
0 & -1
\end{pmatrix},
\qquad
t^2=\frac{1}{2}
\begin{pmatrix}
0 & 1
\\
1 & 0
\end{pmatrix}.
\end{gather}
However, we need to be careful about normalisation.
The normalisation of $\{t^a\}_{a=0,1,2}$ is f\/ixed by the commutation relations~\eqref{commutation relations t's}.
The normalisation of the basis $\{P^{*a}\}_{a=0,1,2}$, which is dual to the basis $\{P_{a}\}_{a=0,1,2}$ used
in~\eqref{def commutation relations Poincare}, may be dif\/ferent.
Therefore, we should allow
\begin{gather*}
P^{*a} = \lambda t^a,
\qquad
a=0,1,2,
\end{gather*}
where $\lambda$ is an arbitrary constant of dimension inverse mass.
In Section~\ref{sec: deforming}, we use the invariant pairing~\eqref{def bilinear form} to identify
$\mathfrak{sl}(2,\mathbb{R})^*$ with $\mathfrak{sl}(2,\mathbb{R})$, and $P^{*a}$ with $8\pi G J^a$.
The commutation relations~\eqref{def commutation relations Poincare} then f\/ix $ \lambda = 8\pi G$.

We denote elements of momentum space $\mathfrak{sl}(2,\mathbb{R})$ as~$p$, which we expand as
\begin{gather*}
p = p_a P^{*a} = \lambda p_a t^a.
\end{gather*}
The adjoint action of ${\rm SL}(2,\mathbb{R})$ on $\mathfrak{sl}(2,\mathbb{R})$ leaves invariant the inner product
\begin{gather}\label{Minkowski product Killing form}
-\frac{2}{\lambda^2} \tr(pq)= p_a q^a.
\end{gather}
In the following, we take $p^2$ to mean $p_ap^a$, not the square of the matrix~$p$.

The orbits of the ${\rm SL}(2,\mathbb{R})$ adjoint action on $p\in \mathfrak{sl}(2,\mathbb{R})$ are labelled by the value of
the invariant inner product $p^2$.
The dif\/ferent cases are naturally distinguished by the timelike (T), spacelike (S) or lightlike (L) nature of the
elements~$p$ on a~given orbit.

  {\bf T}:
There are two disjoint families of orbits, corresponding to the dif\/ferent possible signs of a~real parameter $m\neq 0$.
Starting from the timelike representative element $\hat{p}=\lambda m t^0$, the orbits
\begin{gather*}
O^{T}_m = \big\{v \lambda m t^0 v^{-1} \,\big|\, v \in {\rm SL}(2,\mathbb{R})\big\}=\left\{\lambda p_a t^a \in \mathfrak{sl}(2,\mathbb{R})
\,\Big|\, p^2=m^2,\, \frac{p_0}{m}>0 \right\}
\end{gather*}
are the `forward' and `backward' sheets of the two-sheeted mass hyperboloid for, respectively, $m>0$ and $m<0$.
The associated stabiliser group is
\begin{gather*}
N^{T} = \big\{\exp\big(\phi t^0\big) \, |\, \phi \in [0,4\pi) \big\}\simeq {\rm U}(1).
\end{gather*}
Its UIR's are labelled by $s \in \frac{1}{2}\mathbb{Z}$; the half-integer values arise because of the range of $\phi$
for elements of the form $e^{\phi t^0}\in {\rm SL}(2,\mathbb{R})$.

The parameters $\abs{m}$ and~$s$ can be interpreted as the mass and the spin of a~particle.
We allow~$m$ to be either positive or negative, corresponding to the cases of a~particle or antiparticle.
Further note that, in contrast to the 3+1 dimensional case, the spin~$s$ can also be either positive or negative.
In fact, spin in 2+1 dimensions violates parity~$P$ and time-reversal~$T$ unless two species with opposite spin are
included in a~theory~\cite{SousaGerbert}.

 {\bf S}: Picking a~typical spacelike representative element $\hat{p}=\lambda \mu t^1$, the resulting orbit
\begin{gather*}
O^S_\mu = \big\{v \lambda \mu t^1 v^{-1}\,\big|\, v \in {\rm SL}(2,\mathbb{R})\big\}
=\big\{\lambda p_a t^a \in \mathfrak{sl}(2,\mathbb{R})\,\big|\, p^2=-\mu^2<0\big\}
\end{gather*}
is a~single-sheeted hyperboloid.
The real parameter $\mu$ is strictly positive.
The associated stabiliser is
\begin{gather*}
N^S = \big\{{\pm} \exp\big(\vartheta t^1\big)\,|\, \vartheta \in \mathbb{R} \big\} \simeq \mathbb{R}
\times
\mathbb{Z}_2,
\end{gather*}
and its UIR's are labelled by pairs $(s,\epsilon)$, with $s \in \mathbb{R}$, $\epsilon=\pm1$.
Empirically, particles with spacelike momenta~-- so-called tachyons~-- do not exist in the physical 3+1 dimensions.

  {\bf L}: There are again two possibilities corresponding to the dif\/ferent possible signs of $p_0$.
Picking the lightlike representative elements $\hat{p}= \pm E^+ = \pm (t^0+t^2)$ introduced in~\eqref{def HE+E-},
we obtain the `forward' and `backward' light cones as orbits:
\begin{gather*}
O^{L\pm} = \big\{\pm v E^+ v^{-1}\, |\, v \in {\rm SL}(2,\mathbb{R}) \big\}=\big\{\lambda p_a t^a \in \mathfrak{sl}(2,\mathbb{R})\,|\,
p^2=0, \, \pm p_0>0 \big\}.
\end{gather*}
The stabiliser group in both cases is
\begin{gather*}
N^{L}= \big\{{\pm} \exp(zE^+)\,|\, z \in \mathbb{R}\big\} \simeq \mathbb{R}
\times
\mathbb{Z}_2.
\end{gather*}
Its UIR's are again labelled by pairs $(s,\epsilon)$, with $s \in \mathbb{R}$, $\epsilon=\pm1$.

 {\bf V}: The `vacuum' orbit $\{0\}$ consists solely of the origin and the associated stabiliser is the whole
group ${\rm SL}(2,\mathbb{R})$.
The irreducible representations of ${\rm SL}(2,\mathbb{R})$ can, for instance, be found in~\cite{Knapp}.

There are two standard ways of writing down the UIR's of semidirect product groups like~$\tilde{P}_{3}$, both using the
orbits and stabiliser UIR's listed above.
One uses sections of bundles over the homogeneous space~${\rm SL}(2,\mathbb{R}) / N$, where~$N$ denotes one of the stabiliser groups.
The group action on such sections involves multipliers or cocycles, see~\cite{BR} for details.
The other uses functions on~${\rm SL}(2,\mathbb{R})$ satisfying an equivariance condition.
This is the formulation we use here, referring the reader to~\cite{BMS, BR} for a~translation between the two approaches.

For a~given UIR of $\tilde{P}_{3}$ labelled by an orbit~$O$ with representative element $\hat{p}$, stabiliser group~$N$
and UIR $\varsigma$ of~$N$ on a~vector space~$V$, the carrier space is
\begin{gather}\label{carrier space}
V_{O,\varsigma}= \big\{\psi: {\rm SL}(2,\mathbb{R}) \rightarrow V \,|\, \psi(vn)= \varsigma\big(n^{-1}\big)\psi(v), \, \forall\, n \in N,
\, \forall\,  v \in {\rm SL}(2,\mathbb{R})\big\}.
\end{gather}
We also have to impose an integrability condition, which we give in a~particular case below.
An element $(g,a) \in \tilde{P}_{3}$ acts on $\psi \in V_{O,\varsigma}$ via
\begin{gather*}
\pi_{O,\varsigma}((g,a))\psi(v) = \exp \big(ia(\Ad_{g^{-1}v}(\hat{p}))\big)\psi\big(g^{-1}v\big).
\end{gather*}

As we will subsequently focus on the case of timelike momenta, we give the carrier space for this case explicitly:
\begin{gather}\label{carrier space a)}
V_{ms}= \big\{\psi: {\rm SL}(2,\mathbb{R}) \rightarrow \mathbb{C} \,\big|\, \psi\big(ve^{\alpha t^0}\big)= e^{-i s \alpha}\psi(v), \, \forall
\, (\alpha, v) \in [0,4\pi)\times {\rm SL}(2,\mathbb{R})\big\}.
\end{gather}
The integrability condition is
\begin{gather*}
\int_{{\rm SL}(2,\mathbb{R})/N^T} \lvert \psi \rvert^2(w)\, d\nu(w) < \infty.
\end{gather*}
Here, $d\nu$ is the invariant measure on the homogeneous space ${\rm SL}(2,\mathbb{R})/N^T$ (note that $\lvert \psi \rvert^2$
only depends on $w \in {\rm SL}(2,\mathbb{R})/N^T$).

An element $(g,a) \in \tilde{P}_{3}$ acts on $\psi \in V_{ms}$ via
\begin{gather*}
\pi_{ms}((g,a))\psi(v) = \exp \big(ia\big(\Ad_{g^{-1}v}\big(\lambda mt^0\big)\big)\big)\psi\big(g^{-1}v\big).
\end{gather*}
If we introduce the notation
\begin{gather}\label{p_i def}
p = \lambda m vt^0v^{-1}
\end{gather}
for an orbit element, this further simplif\/ies to
\begin{gather}\label{P3action}
\pi_{ms}((g,a))\psi(v) = e^{ia(\Ad_{g^{-1}}(p))}\psi\big(g^{-1}v\big).
\end{gather}

\subsection{Covariant momentum constraints}\label{covariant wave equations}

In a~f\/ield theory, we are usually looking for wave functions that are def\/ined on momentum or position space and which
transform covariantly under the action of the Poincar\'e group~\cite{BR,Binegar}.
In our conventions, the required transformation behaviour reads
\begin{gather*}
\pi((g,a))\tilde \phi(p)= e^{ia(\Ad_{g^{-1}}(p))} \rho(g)\tilde \phi\big(g^{-1}p\big),
\end{gather*}
where $\rho$ is a~(preferably f\/inite-dimensional) representation of the full group ${\rm SL}(2,\mathbb{R})$.

To obtain a~covariant description, we employ the technique of~\cite{MS}.
In geometric terms, the approach taken there can be described as follows.
The formulation~\eqref{carrier space} def\/ines elements of the carrier space of an UIR as functions on the group obeying
an equivariance condition.
Replacing ${\rm SL}(2,\mathbb{R})$ with a~general Lie group~$G$ and considering a~general stabiliser subgroup~$N$, this is
nothing but the equivariant description of sections of vector bundles over $G/N$.
For $G={\rm SU}(2)$ and $N={\rm U}(1)$, these are the standard Hermitian line bundles over $S^2$.

The trick used in~\cite{MS} is to view the bundles above as subbundles of the trivial bundle $S^2
\times
\mathbb{C}^n$, where $\mathbb{C}^n$ is the standard~$n$-dimensional UIR of ${\rm SU}(2)$.
In that way, sections become ordinary functions $S^2\rightarrow \mathbb{C}^n$ obeying a~linear constraint.
In this construction, the unitarity of the ${\rm SU}(2)$ action on $\mathbb{C}^n$ is essential for obtaining Hermitian line
bundles.
By thinking of $S^2$ as embedded in Euclidean (momentum) 3-space, one arrives at functions $\mathbb{R}^3\rightarrow
\mathbb{C}^n$ obeying linear constraints.
Applying an ordinary Fourier transform then produces functions on Euclidean (position) 3-space obeying a~linear
dif\/ferential equation.

We would like to treat the Lorentzian situation analogously.
However, the standard~$n$-dimensional irreducible representations of ${\rm SL}(2,\mathbb{R})$, reviewed in
Appendix~\ref{conventions}, are not unitary, and therefore the procedure of~\cite{MS} cannot be used to obtain all UIR's
of $\tilde{P}_{3}$.
We shall now show that it can be implemented for the UIR's~\eqref{carrier space a)} labelled by orbits containing
timelike momenta.
In that case, the stabiliser group is the ${\rm U}(1)$ subgroup of ${\rm SL}(2,\mathbb{R})$ generated by $t^0$.

For a~given $\psi$ in~\eqref{carrier space a)}, we def\/ine the maps
\begin{gather*}
\tilde\phi: \  O^{T}_m \rightarrow \mathbb{C}^{2\abs{s}+1}
\end{gather*}
via
\begin{gather}\label{phi pm def}
\tilde\phi(p)=\psi(v)\rho^{\abs{s}}(v)\lvert \abs{s}, s \rangle,
\end{gather}
where~$p$ is related to~$v$ via~\eqref{p_i def}, and the states $|\abs{s},k\rangle$ form the basis~\eqref{eigenbasis} of
the f\/inite-dimensional $\mathfrak{sl}(2,\mathbb{R})$ irreducible representations in which $t^0$ is diagonal.
Clearly,
\begin{gather*}
\rho^{\abs{s}}\big(ve^{\alpha t^0}\big) \lvert \abs{s},s\rangle = \rho^{\abs{s}}(v)\rho^{\abs{s}}\big(e^{\alpha t^0}\big) \lvert
{\abs{s}},s\rangle = \rho^{\abs{s}}(v)e^{i \alpha s}\lvert {\abs{s}},s\rangle.
\end{gather*}
This cancels the phase picked up by $\psi$ under the right-multiplication by $e^{\alpha t^0}$.
Hence, $\tilde\phi$ only depends on $p \in O^{T}_m$, even though both $\rho^{\abs{s}}(v)$ and $\psi$ depend on~$v$.

We now see why this procedure is generally not feasible for UIR's~\eqref{carrier space} labelled by orbits containing
spacelike or lightlike momenta, where the stabiliser groups are generated by spacelike and lightlike generators in
$\mathfrak{sl}(2,\mathbb{R})$.
Under the right-multiplication by $e^{\alpha t^1}$ resp.~$e^{\alpha t^+}$, the elements of~\eqref{carrier space} pick up a~phase that cannot be compensated using one of the
f\/inite-dimensional irreducible representations of~${\rm SL}(2,\mathbb{R})$, as $\rho^{\abs{s}}(t^1)$ has real eigenvalues and
$\rho^{\abs{s}}(t^+)$ has zero as the sole eigenvalue.

Similar restrictions were found in~\cite{Binegar} for the existence of a~f\/inite-dimensional covariant description.
More general covariant descriptions are given in~\cite{Gitman}.
However, these are not obtained directly from the standard UIR's of the Poincar\'e group.
Instead, they are constructed using generalised regular representations.

The maps $\tilde\phi$ def\/ined in~\eqref{phi pm def} satisfy the constraint
\begin{gather}\label{momconstraint m}
\big(i\rho^{\abs{s}}(t^a) p_a + m s\big)\tilde\phi(p)=0,
\end{gather}
as can be seen by writing~\eqref{p_i def} as $p_a t^a= vmt_0v^{-1}$:
\begin{gather*}
\rho^{\abs{s}}(t_a) p_a \tilde\phi(p)=\rho^{\abs{s}}\big(v m t^0 v^{-1}\big)\rho^{\abs{s}}(v)\psi(v) \lvert \abs{s},s\rangle
 =\psi(v)\rho^{\abs{s}}(v) m i s \lvert \abs{s}, s\rangle
 = i m s \tilde\phi(p),
\end{gather*}
as required.
The equation~\eqref{momconstraint m} later becomes one of our wave equations and we refer to it as the spin constraint.

Following the method of~\cite{MS}, we now consider extensions of the function $\tilde\phi$, def\/ined on the Lie algebra
$\mathfrak{sl}(2,\mathbb{R})$.
This will enable us to employ a~standard Fourier transform for switching from momentum to position space.
We embed the timelike orbits $O^{T}_m$ into the Lie algebra $\mathfrak{sl}(2,\mathbb{R})$ and def\/ine
\begin{gather*}
W_{ms} = \big\{\tilde\phi: \mathfrak{sl}(2,\mathbb{R}) \rightarrow
\mathbb{C}^{2\abs{s}+1} \,\big|\, \big(i \rho^{\abs{s}}(t^a)p_a + m s\big)\tilde\phi(p)=0, \, \big(p^2-m^2\big)\tilde\phi(p)=0\big\}.
\end{gather*}
We call the condition
\begin{gather}\label{mass irreducibility condition}
\big(p^2-m^2\big)\tilde\phi(p)=0
\end{gather}
the mass constraint; it is the only condition for spin $s=0$ and we will see that it is implied by the spin constraint
for the cases $s= \pm \frac{1}{2}, \pm 1$.

The spaces $ W_{ms}$ carry a~representation of $\tilde{P}_{3}$ which we shall give below.
However, the mass constraint does not f\/ix the sign of~$m$.
In order to obtain irreducible representations of $\tilde{P}_{3}$, we therefore still need to impose
\begin{gather}\label{sign irreducibility condition}
\Theta\left(-\frac{p_0}{m}\right)\tilde{\phi}(p)=0,
\end{gather}
where $\Theta$ is the Heaviside step function.
We call this condition the sign constraint.
We remark that though~$W_{ms}$ are reducible representations of~$\tilde{P}_{3}$, they are irreducible representations of
a~suitable double cover of the extended Poincar\'{e} group, which includes time reversal (mapping~$O^{T}_{m}$ to~$O^{T}_{-m}$).

The action of an element $(g,a) \in \tilde{P}_{3}$ on $\tilde\phi \in W_{ms}$ is
\begin{gather*}
\big(\pi_{m s}((g,a))\tilde\phi\big) (p)= e^{i a(\Ad_{g^{-1}}p)}\rho^{\abs{s}}(g)\tilde\phi\big(\Ad_{g^{-1}}p\big).
\end{gather*}
It commutes with the constraints~\eqref{momconstraint m},~\eqref{mass irreducibility condition} and~\eqref{sign
irreducibility condition}, as required.

Before we can claim that this is an UIR, we need to def\/ine the inner product with respect to which the representations
are unitary.
For spin $0$, the invariant inner product on the space $W_{m,s=0}$ is the familiar
\begin{gather*}
\big(\tilde\phi_1,\tilde\phi_2\big) = \int_{O^{T}_m\cup O^{T}_{-m}} \tilde\phi^*_1\tilde\phi_2   \frac{dp_1
dp_2}{\abs{p_0}},
\end{gather*}
where the integration is with respect to the standard Lorentz-invariant measure on the mass shell.
We will give the inner product for spin $\pm \frac 1 2 $ and spin $\pm 1$ below.
For a~general discussion of the construction of the required invariant scalar product, see~\cite{BR}.

In the case $s=\frac{1}{2}$, the spin constraint~\eqref{momconstraint m} becomes the Dirac equation in momentum space
\begin{gather}\label{Dirac equation momentum space}
\left(i t^a p_a + \frac{1}{2}m\right)\tilde\phi(p)=0.
\end{gather}
Applying $(i t^a p_a - \frac{1}{2}m)$ to this and using~\eqref{anticommutation relations t's}, we see that~\eqref{Dirac
equation momentum space} implies the mass constraint~\eqref{mass irreducibility condition} but not the sign
constraint~\eqref{sign irreducibility condition}.
However, $\tilde\phi$ can be decomposed into positive and negative frequency parts $\tilde\phi^+$ and $\tilde\phi^-$
using a~Foldy-Wouthuysen transformation; see~\cite{Binegar} for details.
This is completely analogous to the situation in 3+1 dimensions.

To see that~\eqref{Dirac equation momentum space} is indeed the Dirac equation in momentum space, we note that in 2+1
dimensions, Clif\/ford generators (gamma matrices) satisfying
\begin{gather*}
\big\{\gamma^a,\gamma^b\big\}=\gamma^a\gamma^b+\gamma^b\gamma^a= 2\eta^{ab}\id
\end{gather*}
can be obtained from the $\mathfrak{sl}(2,\mathbb{R})$ generators~\eqref{def t_a's} via
\begin{gather}\label{def gammas}
\gamma^a=2i t^a.
\end{gather}
Thus we can write~\eqref{Dirac equation momentum space} as
\begin{gather}\label{Dirac equation momentum space 2}
(\gamma^a p_a + m)\tilde\phi(p)=0.
\end{gather}

The invariant scalar product on the space $W_{m,s=\frac{1}{2}}$ is
\begin{gather*}
\big(\tilde\phi_1,\tilde\phi_2\big)
= \int_{O^{T}_m\cup O^{T}_{-m}} \tilde\phi_1^\dagger \gamma^0\tilde\phi_2\frac{dp_1 dp_2}{\abs{p_0}}.
\end{gather*}
The Lorentz invariance of $\tilde\phi_1^\dagger \gamma^0\tilde\phi_2$ follows from the KAN or Iwasawa decomposition of
an element $g\in {\rm SL}(2,\mathbb{R})$ into $g=kv$, where~$k$ is a~rotation (generated by~$t^0$ and commuting with
$\gamma_0$) and~$v$ is of the form
\begin{gather*}
v=
\begin{pmatrix}
r & x
\\
0 & \frac 1 r
\end{pmatrix},
\qquad
r>0,
\qquad
x\in \mathbb{R}.
\end{gather*}
It satisf\/ies $v^t\gamma^0v= \gamma^0$.

For $s=1$, $\tilde\phi=\tilde\phi_at^a$ takes values in the adjoint representation of $\mathfrak{sl}(2,\mathbb{R})$.
The constraint~\eqref{momconstraint m} then gives the Proca equations in momentum space
\begin{gather*}
\big(i p_a \ad(t^a) + m\big)\tilde\phi(p)=0,
\end{gather*}
or
\begin{gather}\label{Proca equation momentum space 2}
\big[p_a t^a,\tilde\phi(p)\big] = i m \tilde\phi(p).
\end{gather}
Taking the Minkowski product~\eqref{Minkowski product Killing form} with $p_dt^d$ gives
\begin{gather}\label{Proca equation momentum space 3}
p^a\tilde\phi_a(p)=0.
\end{gather}
The previous two equations together with the identity
\begin{gather*}
[\xi,[\eta,\zeta]]= (\xi_a \zeta^a)\eta- (\xi_a\eta^a) \zeta,
\qquad
\xi,\eta,\zeta \in \mathfrak{sl}(2,\mathbb{R}), \qquad \xi=\xi_at^a \quad \text{etc.}
\end{gather*}
give the mass constraint~\eqref{mass irreducibility condition}.
Like for spin $\frac 1 2 $, the equation~\eqref{momconstraint m} implies the mass constraint~\eqref{mass irreducibility
condition} but not the sign constraint~\eqref{sign irreducibility condition}.

The invariant scalar product on the space $W_{m,s=1}$ is
\begin{gather}\label{scalar product s=1 momentum space}
\big(\tilde\phi_1, \tilde\phi_2\big) = -\int_{O^{T}_m\cup O^{T}_{-m}} \tilde\phi^*_{1a}
\tilde\phi^{\phantom{\dagger}a}_{2}  \frac{dp_1 dp_2}{\abs{p_0}}.
\end{gather}
This is manifestly Lorentz invariant, but it may not be obvious that~\eqref{scalar product s=1 momentum space} is indeed
positive def\/inite.
This can be seen as follows: due to~\eqref{Proca equation momentum space 3} $\tilde\phi$ is spacelike, and $\eta$ is
negative def\/inite when restricted to spacelike vectors.

The wave equations for the cases $s=-\frac{1}{2}$ and $s=-1$ can be obtained from~\eqref{Dirac equation momentum space
2} and~\eqref{Proca equation momentum space 2} by changing the sign in front of~$m$, while the respective inner products
stay the same.

\subsection{Fourier transform to position space}\label{Fourier transformation}

The momentum-space form of the UIR's of $\tilde{P}_{3}$ in the previous sections were designed to be amenable to
a~standard Fourier transform.
Def\/ining
\begin{gather*}
\phi(x)= \int e^{i x(p)}\tilde\phi(p)\,d^3p,
\end{gather*}
the spin constraint~\eqref{momconstraint m} turns into the f\/irst order dif\/ferential equation
\begin{gather*}
\big(\rho^{\abs{s}}(t^a) \partial_a + m s\big) \phi(x)=0.
\end{gather*}
The mass constraint~\eqref{mass irreducibility condition} becomes the Klein--Gordon equation
\begin{gather*}
\big(\Box+m^2\big)\phi=0.
\end{gather*}
These are the general wave equations for massive particles with spin $s\in \frac 1 2 \mathbb{Z}$ in 2+1 dimensions.
An element $(g,a) \in \tilde{P}_{3}$ acts on the wave function $\phi$ via
\begin{gather*}
\left(\pi_{m s}((g,a))\phi\right) (x)= \rho^{\abs{s}}(g)\phi\big(\Ad^*_g x + a\big).
\end{gather*}

The wave equations for low values of the spin are some of the most studied equations of relativistic physics.
For spin $0$, the mass constraint is the only constraint, and we obtain the Klein--Gordon equation as already noted
above.
For spin $\frac 1 2 $, the spin constraint~\eqref{Dirac equation momentum space 2} Fourier transforms to the Dirac
equation in position space:
\begin{gather*}
(i \gamma^a \partial_a - m)\phi=0.
\end{gather*}
For spin 1, the condition~\eqref{Proca equation momentum space 2} becomes the Proca equation
\begin{gather*}
\partial_a [ t^a,\phi]= - m \phi,
\end{gather*}
and the constraint~\eqref{Proca equation momentum space 3} becomes
\begin{gather*}
\partial^a\phi_a=0.
\end{gather*}

\section{Deforming momentum space}\label{sec: deforming}

\subsection[The quantum double $\mathcal{D}({\rm SL}(2,\mathbb{R}))$: motivation and def\/inition]
{The quantum double $\boldsymbol{\mathcal{D}({\rm SL}(2,\mathbb{R}))}$: motivation and def\/inition}

{\sloppy We now repeat the analysis in the previous section for the case of the quantum double $\mathcal{D}({\rm SL}(2,\mathbb{R}))$
of ${\rm SL}(2,\mathbb{R})$, or Lorentz double for short.
Before summarising the def\/ining pro\-per\-ties of the quantum double of a~Lie group, we make a~few qualitative remarks which
highlight the relation between the Lorentz double and the Poincar\'e group, following~\cite{BMS,Schroers}.

}

The action~\eqref{P3action} of a~Poincar\'e group element on an element of one of its UIR's shows that pure translations
act by a~multiplication with a~special function on the (linear) momentum space~$\mathfrak{sl}(2,\mathbb{R})$, namely the
plane wave $\psi_a(p)=e^{ia(p)}$.
In the Lorentz double, this is deformed and generalised: the momentum space is exponentiated and extended to become the
whole group mani\-fold~${\rm SL}(2,\mathbb{R})$.
The space of functions on momentum spaces is generalised to a~suitably well-behaved class, for example the class of
continuous functions~\cite{KM}.
This deforms the translation part of the Poincar\'e group into something dual to the rotation/boost part: translations
are functions on~${\rm SL}(2,\mathbb{R})$ and rotations/boost are elements of~${\rm SL}(2,\mathbb{R})$.
By allowing linear combinations we obtain a~Hopf algebra, consisting of two subalgebras which are in duality.

The quantum double of a~Lie group is an example of a~quantum double, which in turn is a~special class of quantum
groups~\cite{Drinfeld, Majidbook}.
It can be def\/ined in various ways.
Here we use the form given in~\cite{BM,KM} for locally compact Lie groups.
As a~vector space, the quantum double~$\mathcal{D}(G)$ of a~Lie group~$G$ is the space of continuous, complex-valued
functions~$C(G
\times
G)$.
Morally, one should think of this as the tensor product $C(G)\otimes C(G)$, with the f\/irst factor being the group
algebra and the second factor being the function algebra on~$G$.
The product in the f\/irst factor is by convolution and the product in the second factor is pointwise, but twisted by the
action of the f\/irst argument.
The identity cannot be written as an element of~$C(G
\times
G)$.
Strictly speaking it should be added as a~separate element, but it is convenient to formally express it as
a~delta-function.

In the conventions of~\cite{MS} (which dif\/fer from those in~\cite{BM,KM}), the product $\bullet$, coproduct $\Delta$,
unit $1$, co-unit $\varepsilon$, antipode~$S$ and $*$-structure are as follows
\begin{gather*}
(F_1\bullet F_2)(g,u) :=\int_G F_1\big(z,zuz^{-1}\big)  F_2\big(z^{-1}g,u\big) \, dz,
\\
1(g,u) :=\delta_e(g),
\\
(\Delta F)(g_1,u_1;g_2,u_2) :=F(g_1,u_1u_2) \delta_{g_1}(g_2),
\\
\varepsilon(F) :=\int_G F(z,e) \, dz,
\\
(S F)(g,u) :=F\big(g^{-1},g^{-1}u^{-1}g\big),
\\
F^*(g,u) :=\overline{F\big(g^{-1},g^{-1}ug\big)}.
\end{gather*}
In these equations, all integrals over the group are with respect to the Haar measure and $e\in G$ denotes the identity element.
The quantum double is quasitriangular~\cite{Drinfeld}, and the expression for the~$R$-matrix can be found in~\cite{BM,KM}.
We do not require it here.

\subsection[Coordinates for ${\rm SL}(2,\mathbb{R})$]{Coordinates for $\boldsymbol{{\rm SL}(2,\mathbb{R})}$}

There are many natural ways to coordinatise the Lie group ${\rm SL}(2,\mathbb{R})$, see~\cite{ALL} for a~recent review in the
context of 2+1 gravity.
Here we use two sets of coordinates, one obtained via the exponential map $\mathfrak{sl}(2,\mathbb{R}) \rightarrow
{\rm SL}(2,\mathbb{R})$ and a~second which exploits the realisation of ${\rm SL}(2,\mathbb{R})$ as a~submanifold of $\mathbb{R}^4$.

The exponential map $\exp:\mathfrak{sl}(2,\mathbb{R})\rightarrow {\rm SL}(2,\mathbb{R})$ is bijective when restricted to
a~suf\/f\/iciently small neighbourhood of $0\in \mathfrak{sl}(2,\mathbb{R}) $ and $\id\in {\rm SL}(2,\mathbb{R})$, but this
is not the case globally.
In fact, it is neither injective nor surjective as we shall see in our discussion of conjugacy classes below.
As before, we write elements of $\mathfrak{sl}(2,\mathbb{R})$ as $p=\lambda p_at^a$.
Using the fact that $\lambda p_a t^a$ squares to $-\frac{\lambda^2}{4} p^2 \id$, one f\/inds:
\begin{gather}\label{exp(slt)}
\exp(\lambda p_at^a)=
\begin{cases}
\cos\big(\lambda \sqrt{p^2}/2\big)\id + \dfrac{p_a}{\sqrt{p^2}/2}\sin\big(\lambda \sqrt{p^2}/2\big)t^a, & \text{if}\quad p^2>0,
\vspace{1mm}\\
\id + \lambda p_a t^a, & \text{if}\quad  p^2=0,
\vspace{1mm}\\
\cosh\big(\lambda \sqrt{-p^2}/2\big)\id + \dfrac{p_a}{\sqrt{-p^2}/2}\sinh\big(\lambda \sqrt{-p^2}/2\big)t^a, & \text{if}\quad p^2<0.
\end{cases}
\end{gather}
It follows from these formulae that elements $u\in {\rm SL}(2,\mathbb{R})$ with $\tr(u)<-2$ cannot be written as
exponentials.
As we shall see in Section~\ref{Induced representations DSLt} below, some elements with $\tr(u)=-2$ can also not be
written as exponentials.
However, we shall also see that if~$u$ is not in the image of the exponential map, then $-u$ is.
This fact will be useful in Section~\ref{noncommutative Fourier transformation}.

To realise ${\rm SL}(2,\mathbb{R})$ as a~submanifold of $\mathbb{R}^4$, we introduce Cartesian coordinates
$(\mathcal{P}_0,\mathcal{P}_1,\mathcal{P}_2,\mathcal{P}_3)$ on $\mathbb{R}^4$ and expand
\begin{gather}\label{def cP}
u=\mathcal{P}_3\id+\lambda\mathcal{P}_at^a=
\begin{pmatrix}
\mathcal{P}_3+\frac{1}{2}\lambda\mathcal{P}_1 & \frac{1}{2}\lambda\mathcal{P}_0+\frac{1}{2}\lambda\mathcal{P}_2
\\[2mm]
-\frac{1}{2}\lambda\mathcal{P}_0+\frac{1}{2}\lambda\mathcal{P}_2 & \mathcal{P}_3-\frac{1}{2}\lambda\mathcal{P}_1
\end{pmatrix}
,
\end{gather}
where Latin indices still take values $0$, $1$, $2$.
The condition $u\in {\rm SL}(2,\mathbb{R})$ is then equivalent to
\begin{gather}\label{cP squared summed equals one}
\det u = \mathcal{P}_3^2 + \frac{\lambda^2}{4}\mathcal{P}^a\mathcal{P}_a = 1.
\end{gather}
We regard $\mathcal{P}_a$, $a=0,1,2$, as the independent coordinates with
$\mathcal{P}_3=\pm\sqrt{1-\frac{\lambda^2}{4}\mathcal{P}^a\mathcal{P}_a}$.
In the following, we refer to the subsets of ${\rm SL}(2,\mathbb{R})$ with $\mathcal{P}_3\gtrless0$ as the upper and lower
half of ${\rm SL}(2,\mathbb{R})$.

Comparing~\eqref{exp(slt)} and~\eqref{def cP}, we can easily write down a~relation between the two coordinate systems on
the intersections of their respective patches.
The case $p^2>0$ is particularly important for us.
Here one has
\begin{gather*}
\mathcal{P}_3=\cos\big(\lambda\sqrt{p^2}/2\big),
\qquad
\mathcal{P}_a=p_a\frac{\sin\big(\lambda\sqrt{p^2}/2\big)}{\lambda\sqrt{p^2}/2}.
\end{gather*}

Taking the limit $\lambda \rightarrow 0$ corresponds to the f\/lattening out of momentum space ${\rm SL}(2,\mathbb{R})={\rm AdS}_3$.
It f\/inally rips apart in the hyperplane of $\mathcal{P}_3=0$, producing not one but two copies of f\/lat Minkowski
momentum space situated at $\mathcal{P}_3=\pm1$.
They would be identif\/ied if we had worked with $L_3^{+\uparrow}$ instead of ${\rm SL}(2,\mathbb{R})$.
If, on the other hand, we had worked with the universal covering group, we would have found a~countable inf\/inity of
copies.
For a~discussion of $L_3^{+\uparrow}$ as momentum space in (2+1)-dimensional gravity and (2+1)-dimensional
non-commutative scalar f\/ield theories, see~\cite{SS}.

This property of momentum space is an important consequence of the transition to the double cover or universal cover of
$P_3$, compounding the more widely known manifestation via the spin of massive particles, which takes integer values in
the case of $P_3$, half-integer values in the case of $\tilde{P}_{3}$ and real values in the case of the universal cover
of $P_3$ (see our discussion in Section~\ref{Poincare group}).

\subsection[Irreducible representations of $\mathcal{D}({\rm SL}(2,\mathbb{R}))$]{Irreducible representations of $\boldsymbol{\mathcal{D}({\rm SL}(2,\mathbb{R}))}$}
\label{Induced representations DSLt}

The Lorentz double $\mathcal{D}({\rm SL}(2,\mathbb{R}))$ is a~special example of a~transformation group algebra, and its UIR's
can best be understood in that general context.
As shown in~\cite{KM}, they are labelled by conjugacy classes in ${\rm SL}(2,\mathbb{R})$ and UIR's of the associated
centraliser or stabiliser groups.
As emphasised in~\cite{BMS,Schroers}, this should be seen as a~deformation of the picture for the semi-direct product
group $\tilde{P}_{3}$.
In both cases, the UIR's are labelled by ${\rm SL}(2,\mathbb{R})$ orbits in momentum space and UIR's of associated
stabilisers.
The dif\/ference is that momentum space is linear for $\tilde{P}_{3}$ and curved for $\mathcal{D}({\rm SL}(2,\mathbb{R}))$.

The conjugacy classes of ${\rm SL}(2,\mathbb{R})$ and their associated stabilisers are classif\/ied in~\cite{KM}, and we list
them here in a~notation adapted to our needs.
From the def\/ining property of ${\rm SL}(2,\mathbb{R})=\{g\in {\rm GL}(2,\mathbb{R}) \,|\, \det(g)=1\}$ it follows that the
(generalised) eigenvalues $\lambda_1$, $\lambda_2$ of a~given element multiply to one.
They are thus either complex conjugate to each other or both real.
The set of conjugacy classes can be organised according to the dif\/ferent possible eigenvalues.
Some but not all of the conjugacy classes can be obtained from the adjoint orbits in the Lie algebra~$\mathfrak{sl}(2,\mathbb{R})$ by exponentiation.
We have chosen a~labelling of the conjugacy classes which mimicks the conventions we used for the adjoint orbits in the
Lie algebra: we use the superscripts~T,~S and~L for `timelike', `spacelike' and `lightlike' to denote conjugacy classes
whose elements can be obtained via exponentiated timelike, spacelike or lightlike elements of
$\mathfrak{sl}(2,\mathbb{R})$.
Our list also includes the stabiliser group of a~representative element in each of the conjugacy classes.

 {\bf T}: For $\lambda_1=e^{i \frac{\theta}{2}}$, $\lambda_2=e^{-i \frac{\theta}{2}}$ $(0<\theta<2\pi)$, there
are two disjoint families of conjugacy classes, with representative elements $\hat{h}= \exp(\pm \theta t^0)$ which are
exponentials of timelike $\mathfrak{sl}(2,\mathbb{R})$ elements.
As for the Lie algebra orbits, we introduce a~unif\/ied notation for the two families, with $\theta \in (0,2\pi)$ to
parametrise one component and $\theta \in (-2\pi,0)$ to parametrise the other:
\begin{gather}\label{timecc}
C^{T}(\theta) =\big\{v\exp\big(\theta t^0\big) v^{-1} \,|\, v \in {\rm SL}(2,\mathbb{R})\big\},
\qquad
\theta \in (-2\pi,0)\cup (0,2\pi).
\end{gather}
The stabiliser group is
\begin{gather*}
N^T = \{\exp\big(\phi t^0\big)\,|\, \phi \in [0,4\pi) \} \simeq {\rm U}(1),
\end{gather*}
with UIR's labelled by $s \in \frac 1 2 \mathbb{Z}$.

 {\bf S}: There is one family of conjugacy classes with eigenvalues of the form $\lambda_1=e^{\frac{r}{2}}$,
$\lambda_2=e^{-\frac{r}{2}}$ $(r \in \mathbb{R}_+)$.
Elements of a~given conjugacy class are obtained by exponentiating a~spacelike Lie algebra element:
\begin{gather*}
C^S(r)=\big\{v\exp\big(rt^1\big) v^{-1} \,\big|\, v \in {\rm SL}(2,\mathbb{R})\big\},
\qquad
r \in \mathbb{R}_+.
\end{gather*}
It has stabiliser group
\begin{gather*}
N^S =\big\{\pm \exp\big(\vartheta t^1\big)\,|\, \vartheta \in \mathbb{R}\big\} \simeq \mathbb{R}
\times
\mathbb{Z}_2,
\end{gather*}
with UIR's labelled by pairs $(b,\epsilon)$, with $b \in \mathbb{R}$, $\epsilon=\pm 1$.

 {\bf -S}: For $\lambda_1=-e^{\frac r 2}$, $\lambda_2= -e^{-\frac r 2}$ $(r \in \mathbb{R}_+)$, there is
likewise one family of conjugacy classes which we write as $-C^{S}(r)$.
Elements are obtained from those of $C^{S}(r)$ by multiplication with $-\id$; they cannot be written as the
exponential of a~Lie algebra element.
The stabiliser group is again $N^S$.

 {\bf L}, {\bf V}: For $\lambda_1=\lambda_2=1$, we distinguish three conjugacy classes: $C^V$, $C^{L+}$ and $C^{L-}$.
The `vacuum' conjugacy class $C^V=\{\id\}$ has stabiliser ${\rm SL}(2,\mathbb{R})$, whose UIR's are discussed
in~\cite{Knapp}.
The lightlike conjugacy classes have representative elements $\hat{h}=\exp(\pm E_+)$, which are the exponentials of the
lightlike elements $\pm E_+$:
\begin{gather*}
C^{L\pm}=\big\{v\exp(\pm E_+)v^{-1} \,\big|\, v \in {\rm SL}(2,\mathbb{R})\big\}.
\end{gather*}
The stabiliser group in both cases is{\samepage
\begin{gather*}
N^{L}=\{\pm \exp(z E_+)\,|\, z\in\mathbb{R} \} \simeq \mathbb{R}
\times
\mathbb{Z}_2,
\end{gather*}
with UIR's labelled by pairs $(b,\epsilon)$, with $b \in \mathbb{R}$, $\epsilon=\pm1$.}

{\bf -L}, {\bf -V}: For $\lambda_1=\lambda_2=-1$, we distinguish three conjugacy classes, which are obtained
by multiplying $C^V$, $C^{L+}$ and $C^{L-}$ by $-\id$.
They have the same stabiliser groups as~$C^V$,~$C^{L+}$ and~$C^{L-}$.
Elements of~$-C^{L+}$ and~$-C^{L-}$ cannot be obtained by exponentiation.

The carrier spaces of the irreducible representations of $\mathcal{D}({\rm SL}(2,\mathbb{R}))$, discussed in~\cite{KM}, are
again given in terms of functions on ${\rm SL}(2,\mathbb{R})$ satisfying an equiva\-rian\-ce condition.
The equiva\-rian\-ce condition only depends on the stabiliser group of a~given conjugacy class, but not directly on the
conjugacy class.
Since the same stabiliser groups arise for orbits in $\mathfrak{sl}(2,\mathbb{R})$ as for conjugacy classes in
${\rm SL}(2,\mathbb{R})$, the general form of the carrier spaces~\eqref{carrier space} of UIR's of $\tilde{P}_{3}$ is
unchanged when replacing $\tilde{P}_{3}$ by $\mathcal{D}({\rm SL}(2,\mathbb{R}))$.
However, the action of the elements of $\mathcal{D}({\rm SL}(2,\mathbb{R}))$ is dif\/ferent, and does depend on the conjugacy
class labelling the representation.

Since we are only able to give covariant forms of momentum constraints in the case of massive particles, i.e.,\ timelike
momenta, we restrict ourselves to the corresponding irreducible representations of $\mathcal{D}({\rm SL}(2,\mathbb{R}))$.
The relevant conjugacy classes are the conjugacy classes $C^{T}(\theta)$ given in~\eqref{timecc}.
Motivated by the application of the Lorentz double to quantum gravity in 2+1 dimensions, we identify the angle $\theta$
labelling the conjugacy classes with the mass of a~particle via $\theta= \lambda m$.
This results in a~bounded mass, which is a~well-known feature of (2+1)-dimensional gravity, where $8\pi G m$ determines
a~def\/icit angle in the conical geometry surrounding a~particle of mass~$m$~\cite{BMS,SchroersCracow}.

Summing up, the irreducible representations of the Lorentz double associated with massive particles are labelled by
a~mass parameter
\begin{gather*}
m\in \left(- \frac{2\pi}{\lambda},0 \right) \cup \left(0, \frac{2\pi}{\lambda}\right)
\end{gather*}
and the spin parameter $s\in \frac 1 2 \mathbb{Z}$.
With the carrier space $V_{ms}$ as def\/ined in~\eqref{carrier space a)}, an element $F \in \mathcal{D}({\rm SL}(2,\mathbb{R}))$
acts on $\psi \in V_{ms}$ as
\begin{gather*}
\left(\Pi_{ms}(F)\psi\right)(v) = \int_{{\rm SL}(2,\mathbb{R})} F\big(z,z^{-1}ve^{m \lambda
t^0}v^{-1}z\big)\psi\big(z^{-1}v\big)\, dz,
\end{gather*}
where we again used the conventions of~\cite{MS}.
In the next section, we adapt the covariantisation procedure of Section~\ref{covariant wave equations} to this representation.

\subsection{Deformed covariant constraints}\label{deformed wave equations}

As in Section~\ref{covariant wave equations}, we begin by trading the equivariant function $\psi \in V_{m s}$ for a~map
\begin{gather*}
\tilde\phi: \ C^{T}(\lambda m) \rightarrow \mathbb{C}^{2|s|+1}
\end{gather*}
via
\begin{gather*}
\tilde\phi(u)=\psi(v)\rho^{\abs{s}}(v)\lvert \abs{s}, s \rangle,
\end{gather*}
where the states $|\abs{s},k\rangle$ are again elements of the basis~\eqref{eigenbasis} and $u=ve^{m\lambda
t^0}v^{-1}\in C^{T}(\lambda m)$.

These functions satisfy the analogue of the spin constraint~\eqref{momconstraint m},
\begin{gather}\label{group momconstraint m}
\big(\rho^{\abs{s}}(u) - e^{i\lambda m s}\big)\tilde\phi(u)=0.
\end{gather}
This can be shown by a~short calculation which is entirely analogous to that following~\eqref{momconstraint m}.
Note that this is a~rather natural condition: the value of the function $\tilde\phi$ at~$u$ lies in the eigenspace of
$\rho^{\abs{s}}(u)$ with eigenvalue $e^{i\lambda m s}$.

We now embed the conjugacy classes $C^{T}(\lambda m) $ into the group ${\rm SL}(2,\mathbb{R})$.
They are characterised~by
\begin{gather}\label{Pmass}
\mathcal{P}_3=\cos\left (\frac{\lambda m}{2}\right),
\qquad
\frac{\mathcal{P}_0}{m} >0.
\end{gather}
In analogy to the conditions~\eqref{mass irreducibility condition} and~\eqref{sign irreducibility condition}, we refer
to the f\/irst of these equations as the mass constraint and to the second as the sign constraint.
In terms of~$u$, the mass constraint is
\begin{gather}\label{mass irreducibility condition group}
\left(\frac12\tr(u)-\cos\left(\frac {\lambda m}{2}\right)\right)\tilde\phi(u)=0.
\end{gather}

We thus def\/ine the carrier spaces
\begin{gather}
\tilde W_{ms} = \Bigg\{\tilde\phi: {\rm SL}(2,\mathbb{R})
\rightarrow \mathbb{C}^{2\abs{s}+1} \,\Bigg|\,\big(\rho^{\abs{s}}(u) - e^{im\lambda s}\big)\tilde\phi(u)=0,
\nonumber
\\
\phantom{\tilde W_{ms} = \Bigg\{}
\left(\frac 1 2 \tr(u)-\cos\left(\frac {\lambda m}{2}\right)\right)\tilde\phi(u)=0\Bigg\},
\label{def W ms group}
\end{gather}
and, as in the undeformed case, we will f\/ind that the mass constraint is actually implied by the spin constraint for
spin $\pm\frac 12$ and spin $\pm 1$.
An element $F \in \mathcal{D}({\rm SL}(2,\mathbb{R}))$ acts on $\tilde\phi \in \tilde W_{ms}$ according to
\begin{gather*}
\big(\Pi_{ms}(F)\tilde\phi\big)(u) = \int_{{\rm SL}(2,\mathbb{R})} F\big(z,z^{-1}uz\big)\rho^{\abs{s}}(z) \tilde\phi\big(z^{-1}uz\big)\, dz.
\end{gather*}

For spinless particles, the covariant description involves a~function $\tilde\phi:{\rm SL}(2,\mathbb{R}) \rightarrow
\mathbb{C}$.
The spin constraint is empty, and we only have the mass constraint~\eqref{mass irreducibility condition group}.
Writing it in terms of $\mathcal{P}_3$ as in~\eqref{Pmass} and applying~\eqref{cP squared summed equals one}, we arrive
at
\begin{gather}\label{noncommutative Klein Gordon equation momentum space}
\mathcal{P}_a\mathcal{P}^a \tilde\phi=\left(\frac{\sin(m\lambda/2)}{\lambda/2}\right)^2 \tilde\phi.
\end{gather}
This is our deformed Klein--Gordon equation in momentum space.

In the case $s=\frac{1}{2}$, we have functions $\tilde\phi:{\rm SL}(2,\mathbb{R}) \rightarrow \mathbb{C}^2$ and the
constraint~\eqref{group momconstraint m} becomes simply
\begin{gather}\label{group constraint spin 1/2}
u\tilde\phi(u) = e^{\frac{i}{2}\lambda m}\tilde\phi(u).
\end{gather}
Inserting $u=\mathcal{P}_3\id+\lambda\mathcal{P}_at^a$, this is equivalent to
\begin{gather}\label{step}
\lambda\mathcal{P}_at^a\tilde\phi(u)=\big(e^{\frac{i}{2}\lambda m}-\mathcal{P}_3\big)\tilde\phi(u).
\end{gather}
However, since the vector $({P}_0,{P}_1,{P}_2)$ (like $(p_0,p_1,p_2)$) is timelike in the case under consideration, the
Lie algebra element $\mathcal{P}_at^a$ is conjugate to a~rotation and has imaginary eigenvalues.
Expanding $e^{\frac{i}{2}\lambda m}= \cos(\lambda m / 2) + i \sin(\lambda m / 2)$, the real part of~\eqref{step} is the
promised mass constraint $\mathcal{P}_3 \tilde\phi = \cos(\lambda m / 2)\tilde\phi$, while the imaginary part is
\begin{gather}\label{noncommutative Dirac momentum space}
\left(i \mathcal{P}_at^a + \frac{1}{2} \frac{\sin(\lambda m / 2)}{\lambda/2}\right)\tilde\phi(u)=0.
\end{gather}

This is our deformed Dirac equation in momentum space.
Using~\eqref{def gammas} to write it in terms of $\gamma$-matrices, we f\/ind
\begin{gather}\label{noncommutative Dirac momentum space in gammas}
\left(\mathcal{P}_a\gamma^a + \frac{\sin(\lambda m / 2)}{\lambda/2}\right)\tilde\phi(u)=0.
\end{gather}
Applying $\big(\mathcal{P}_a\gamma^a - \frac{\sin(\lambda m / 2)}{\lambda/2}\big)$ to~\eqref{noncommutative Dirac momentum
space in gammas} gives $\mathcal{P}_a\mathcal{P}^a\tilde\phi=\frac{\sin^2(\lambda m / 2)}{\lambda^2/4}\tilde\phi$, which
is equivalent to the squared version of the mass constraint.
Note that the information whether $\tilde{\phi}$ has support on the upper or lower half of ${\rm SL}(2,\mathbb{R})$ is not
 contained in the spin constraint.

For $s=1$, we again work with the adjoint representation of ${\rm SL}(2,\mathbb{R})$ and think of $\tilde\phi$ as a~map
$\tilde\phi: {\rm SL}(2,\mathbb{R})\rightarrow \mathfrak{sl}(2,\mathbb{R})$, so we can expand $\tilde\phi = \tilde\phi_at^a$.
Hence, the constraint~\eqref{group momconstraint m} becomes
\begin{gather*}
u\tilde\phi(u)u^{-1}=e^{i \lambda m}\tilde\phi(u).
\end{gather*}
Expanding again $u=\mathcal{P}_3\id+\lambda\mathcal{P}_at^a$, and using the `quaternionic' multiplication
rule~\eqref{ta tb} of the generators $t^a$, we deduce
\begin{gather}\label{step1 in group s=1}
\lambda \mathcal{P}_3\big[\mathcal{P}_at^a, \tilde\phi\big]-\frac{\lambda^2}{2}\big(\mathcal{P}_a\mathcal{P}^a\big)\tilde\phi
+\frac{\lambda^2}{2}\big(\mathcal{P}^a\tilde\phi_a\big)\big(\mathcal{P}_bt^b\big)=\big(e^{i\lambda m}-1\big)\tilde\phi,
\end{gather}
where the evaluation at~$u$ is understood everywhere.
Taking the Minkowski product~\eqref{Minkowski product Killing form} with~$\mathcal{P}_bt^b$ and using that $(e^{i\lambda m}-1)\neq 0$,
we conclude that
\begin{gather*}
\mathcal{P}^a\tilde\phi_a=0.
\end{gather*}
Inserting this into~\eqref{step1 in group s=1} and applying~\eqref{cP squared summed equals one} yields
\begin{gather}\label{step2 in group s=1}
\lambda \mathcal{P}_3\big[\mathcal{P}_at^a, \tilde\phi\big]=\big(e^{i \lambda m}+1-2\mathcal{P}_3^2\big)\tilde\phi.
\end{gather}
Again we can argue from the representation theory of $\mathfrak{sl}(2,\mathbb{R})$ reviewed in
Appendix~\ref{conventions} that the eigenvalues of $[\mathcal{P}_at^a, \cdot]$ are imaginary.
With $\mathcal{P}_3$ real and non-vanishing, we deduce that
\begin{gather*}
\big(\cos(\lambda m) +1 - 2\mathcal{P}_3^2\big)\tilde\phi=0,
\end{gather*}
which is the squared mass constraint
\begin{gather*}
\left(\mathcal{P}^2_3 -\cos^2\frac{\lambda m}{2}\right)\tilde\phi=0.
\end{gather*}
Inserting $\mathcal{P}_3 = \cos\frac{\lambda m}{2}$ into~\eqref{step2 in group s=1}, we f\/inally arrive at
\begin{gather}\label{noncommutative s=1 equation momentum space}
- i [\mathcal{P}_at^a, \tilde\phi] =\frac{\sin(\lambda m/2)}{\lambda/2}\tilde\phi.
\end{gather}
This is the deformed Proca equation in momentum space.

The wave equations in momentum space for the cases $s=-\frac{1}{2}$ and $s=-1$ can again be obtained by changing the
sign in front of~$m$ in~\eqref{noncommutative Dirac momentum space in gammas} and~\eqref{noncommutative s=1 equation
momentum space}.

\section{Towards non-commutative wave equations}\label{noncommutative Fourier transformation}

\subsection{General remarks}

The ordinary Fourier transform, as used in Section~\ref{Fourier transformation}, takes the abelian algebra of functions
on a~vector space (in our case, momentum space) to the abelian algebra of functions on its dual (in our case, position space).
It establishes the link between the UIR's of the Poincar\'e group and the fundamental wave equations of free,
relativistic quantum theory.

Having written some of the irreducible representations of the Lorentz double in terms of $\mathbb{C}^n$-valued functions
on the deformed momentum space ${\rm SL}(2,\mathbb{R})$ obeying Lorentz-covariant constraints, we would now like to use
a~suitable Fourier transform to obtain wave equations in the deformed setting.
Our treatment here will be sketchier than in the previous sections, designed to give an overview and to lay the
foundation for a~future, mathematically more complete treatment.
We consider two kinds of Fourier transform.

One version, called quantum group Fourier transform in the following, takes elements of a~given Hopf algebra to elements
of its dual Hopf algebra~\cite{KeMa, Majidbook}; it is def\/ined in a~rather general Hopf-algebraic setting and can, in
particular, be applied to the Hopf algebra of functions on a~Lie group.

A second version maps functions on a~Lie group~$G$ to functions on the dual of the Lie algebra~$\mathfrak{g}^*$,
equipped with a~$\star$-product.
This is studied in dif\/ferent guises in~\cite{DGL, FL2, FM, JMN, SS2} for the case of~$G$ being the rotation group in
three dimensions (or its cover).
It is investigated in a~more general setting of Lie groups satisfying certain technical requirements in~\cite{GOR, Raasakka}.
We call it group Fourier transform in the following\footnote{The name `non-commutative Fourier transform' is also
frequently used in the literature, but since non-commutativity is also a~feature of the quantum group Fourier transform
we prefer the name `group Fourier transform' here.}. The paper~\cite{FM} also includes a~discussion of the relation
between these two kinds of Fourier transforms.

\subsection{Quantum group Fourier transform}

In our deformed theory, momentum space is ${\rm SL}(2,\mathbb{R})$ and the `algebra of momenta' is the algebra
$C({\rm SL}(2,\mathbb{R}))$ of (suitably well-behaved) functions on ${\rm SL}(2,\mathbb{R})$, with pointwise multiplication.
This is a~commutative but not co-commutative algebra.
The quantum group Fourier transform maps elements of this algebra to elements of the dual `position algebra', which can
be taken to be a~suitable class of functions on ${\rm SL}(2,\mathbb{R})$ with multiplication given by convolution (i.e.,\
a~suitable version of the group algebra) or the universal enveloping algebra ${\rm U}(\mathfrak{sl}(2,\mathbb{R}))$, with
generators
\begin{gather*}
\hat{x}^a=i\lambda t^a
\end{gather*}
satisfying the $\mathfrak{sl}(2,\mathbb{R})$ commutation relations
\begin{gather*}
\big[\hat{x}^a,\hat{x}^b\big] = i \lambda \epsilon^{abc}\hat{x}_c.
\end{gather*}
Note that this non-commutative `spin spacetime' has a~long history in the literature of (2+1)-dimensional quantum
gravity, see for example the papers~\cite{FL1, MW,tHooft}.
It is naturally accommodated in the framework of the Lorentz double, for which, in the terminology of~\cite{BaMa},
${\rm U}(\mathfrak{sl}(2,\mathbb{R}))$ is the `Schr\"odinger representation'.

In~\cite{BaMa}, the authors consider the Euclidean situation ${\rm U}(\mathfrak{su}(2))$, and go on to develop a~bi-covariant
calculus on ${\rm U}(\mathfrak{su}(2))$ and to study the quantum group Fourier transform in this case.
This was used in~\cite{MS} to derive non-commutative linear dif\/ferential equations characterising irreducible
representations of the double $\mathcal{D}({\rm SU}(2))$.
We will now show how most of these results can be adapted, at least formally, to the Lorentzian setting.

The required quantum group Fourier transform is a~map from a~suitable class of functions $C({\rm SL}(2,\mathbb{R}))$ to
a~suitable closure of $ {\rm U}(\mathfrak{sl}(2,\mathbb{R}))$.
This closure should include group elements $u\in {\rm SL}(2,\mathbb{R})$, viewed as inf\/inite power series in
${\rm U}(\mathfrak{sl}(2,\mathbb{R}))$.
The fact that the exponential map is not surjective for ${\rm SL}(2,\mathbb{R})$ does not pose any dif\/f\/iculties here since all
elements of ${\rm SL}(2,\mathbb{R})$ can be written as $\pm \exp(\xi)$ for some $\xi\in \mathfrak{sl}(2,\mathbb{R})$, see our
classif\/ication of conjugacy classes in Section~\ref{Induced representations DSLt}.
In order to accommodate the $\mathbb{C}^n$-valued functions in the carrier space~\eqref{def W ms group}, we tensor both
$C({\rm SL}(2,\mathbb{R}))$ and $ {\rm U}(\mathfrak{sl}(2,\mathbb{R}))$ with $\mathbb{C}^n$.

The `plane waves' used in this quantum group Fourier transforms are simply the group elements of ${\rm SL}(2,\mathbb{R})$,
viewed as functions of the non-commutative position vector $\hat{x}= (\hat{x}^0,\hat{x}^1,\hat{x^2})$ according to
\begin{gather*}
\psi(u;\hat{x}):=u= \pm \exp(-i p_a\hat{x}^a)=\pm \exp(\lambda p_at^a) \in {\rm SL}(2,\mathbb{R}).
\end{gather*}
The quantum group Fourier transform of
\begin{gather}\label{momphi}
\tilde\phi: \ {\rm SL}(2,\mathbb{R}) \rightarrow \mathbb{C}^n
\end{gather}
is then
\begin{gather}\label{Fourier}
\phi(\hat{x}) = \int_{{\rm SL}(2,\mathbb{R})} du\, \psi(u;\hat{x}) \tilde\phi\left(u\right),
\end{gather}
where $du$ is the Haar measure on ${\rm SL}(2,\mathbb{R})$.
The expression~\eqref{Fourier} is formal and the analogue of the corresponding expressions for the Euclidean version
used in~\cite{FM}.
Even in that context, it has not been def\/ined in a~mathematically rigorous fashion.

{\sloppy Adapting the bi-covariant calculus developed in~\cite{BaMa} to the Lorentzian `spin spacetime'
${\rm U}(\mathfrak{sl}(2,\mathbb{R}))$ requires a~vector space on which ${\rm U}(\mathfrak{sl}(2,\mathbb{R}))$ acts from both the
left and the right (i.e.,\ a~bimodule).
As for ${\rm U}(\mathfrak{su}(2))$, we can use the space $M_2(\mathbb{C})$ of complex $2\times2$ matrices
on which the generators $t_a$ of ${\rm U}(\mathfrak{sl}(2,\mathbb{R}))$ act via left- and right-multiplication in
the fundamental representation~\eqref{fundrep}.
Dif\/ferential 1-forms are elements of $M_2(\mathbb{C})\otimes {\rm U}(\mathfrak{sl}(2,\mathbb{R}))$, and the Lorentzian version
of the four-dimensional calculus developed in~\cite{BaMa} gives the exterior derivative of group-like elements as
\begin{gather*}
\text{d} u = \frac 1 \lambda (u-\id)\otimes u,
\end{gather*}
where id is the $2\times2$ identity matrix.
Partial derivatives can be computed by expanding the right-hand side in the basis
\begin{gather*}
e_3=\id,
\qquad
e_a= -i t_a,
\qquad
a=0,1,2.
\end{gather*}
In our coordinates~\eqref{def cP}, we have $u=\mathcal{P}_3 \id +\lambda\mathcal{P}_at^a$ and f\/ind
\begin{gather*}
\partial_3 u= \frac 1 \lambda (\mathcal{P}_3-1) u,
\qquad
\partial_a u= i \mathcal{P}_a u,
\qquad
a=0,1,2.
\end{gather*}

}

Assuming the validity of the Fourier transform~\eqref{Fourier}, non-commutative wave equations can now easily be
obtained from our momentum constraints in Section~\ref{deformed wave equations}.
The constraint~\eqref{noncommutative Klein Gordon equation momentum space} implies the non-commutative Klein--Gordon
equation
\begin{gather}\label{noncommutative Klein Gordon equation position space}
\left(\partial_a\partial^a + \left(\frac{\sin(m\lambda/2)}{\lambda/2}\right)^2 \right) \phi =0.
\end{gather}
The deformed spin $\frac 1 2 $ constraint~\eqref{noncommutative Dirac momentum space in gammas} takes the from of
a~non-commutative Dirac equation
\begin{gather}\label{noncommutative Dirac position space}
\left(i \partial_a\gamma^a - \frac{\sin(\lambda m / 2)}{\lambda/2}\right)\phi=0,
\end{gather}
and the deformed Proca constraint~\eqref{noncommutative s=1 equation momentum space} turns into the non-commutative
Proca equation
\begin{gather}\label{noncommutative s=1 equation position space}
\partial_a[t^a,\phi] = - \frac{\sin(\lambda m/2)}{\lambda/2}\phi,
\end{gather}
which implies $ \partial^a\phi_a=0$.

\subsection{Group Fourier transform}

We turn to the group Fourier transform of functions of the form~\eqref{momphi}.
This time, the image of the Fourier transform is a~certain class of function on ordinary $\mathbb{R}^3$, equipped with
a~$\star$-product.
As in our discussion of the quantum group Fourier transform, we will sketch the main ideas here, leaving a~careful
treatment for future work.
Our main references are~\cite{DGL, FL2, FM, JMN, SS2} which deal with the case of ${\rm SU}(2)$ and ${\rm SO}(3)$,
and~\cite{GOR, Raasakka} for a~more general discussion of the group Fourier transform.
These papers discuss dif\/ferent possibilities of implementing the group Fourier transform and the associated
$\star$-products.
The starting point for each possibility is a~choice of plane wave, which, for a~general Lie group~$G$, is a~map
\begin{gather*}
\psi_\star: \  G
\times
\mathfrak{g}^* \rightarrow \mathbb{C},
\end{gather*}
satisfying a~completeness condition~\cite{Raasakka}
\begin{gather*}
\int_{\mathfrak{g}^*} dx\, \psi(u,x)=\delta_e(u),
\end{gather*}
where $dx$ is a~(suitably normalised) measure on the vector space~$\mathfrak{g}^*$, and $\delta_e$ is the Dirac delta
distribution at the identity element $e\in G$.
Evaluating the plane wave on a~given $u\in G$ produces functions on~$\mathfrak{g}^*$ for which we def\/ine
a~$\star$-product via
\begin{gather}\label{starprod}
\psi_\star\big(u^{(1)},x\big)\star \psi_\star\big(u^{(2)},x\big)=\psi_\star\big(u^{(1)}u^{(2)},x\big).
\end{gather}
This induces a~$\star$-product on the space $L^2_\star(\mathfrak{g}^*)$ of all functions on~$\mathfrak{g}^*$ which can
be written as the group Fourier transform of some $\tilde\phi \in C(G)$:
\begin{gather*}
\phi(x) = \int_G du \, \psi_\star(u,x)   \tilde\phi(u).
\end{gather*}

Even for the most studied case $G={\rm SU}(2)$, $\mathfrak{g}^*\simeq \mathbb{R}^3$, it is not easy to write down a~suitable
plane wave.
Several options have been considered in the literature, each with its own advantages and drawbacks.

In~\cite{FL2}, the authors study a~plane wave which is def\/ined for the quotient ${\rm SO}(3)\simeq {\rm SU}(2)/\mathbb{Z}_2$.
This is reviewed in~\cite{FM}, where the authors then go on to treat the case $G={\rm SU}(2)$ by extending it centrally to
$\mathbb{R}^+
\times
{\rm SU}(2)$.
In~\cite{DGL}, plane waves for $G={\rm SU}(2)$ are constructed by using a~spinorial parametrisation of $\mathbb{R}^3$
(essentially by using $x\in \mathbb{R}^3$ to parametrise a~projection operator onto~${\rm SU}(2)$ eigenstates) while
in~\cite{GOR} plane waves for~${\rm SU}(2)$ are constructed using a~parametrisation via the exponential map.

All these constructions can be adapted with dif\/ferent degrees of completeness to the case at hand, i.e.,
$G={\rm SL}(2,\mathbb{R})$, $\mathfrak{g}^*\simeq \mathbb{R}^3$.
One way to bring out the similarities is to view ${\rm SU}(2)$ as the group of unit quaternions and ${\rm SL}(2,\mathbb{R})$ as the
group of unit pseudo-quaternions, see~\cite{MSquat} for a~detailed discussion of this point of view in the context of
2+1 gravity.
The parametrisation~\eqref{def cP} of $u\in {\rm SL}(2,\mathbb{R})$ is essentially a~quaternionic parametrisation, with
$\lambda t_a$ playing the role of imaginary pseudo-quaternions and id being the identity in the pseudo-quaternions.
Then the central extensions $\mathbb{R}^+
\times
{\rm SU}(2)$ and $\mathbb{R}^+
\times
{\rm SL}(2,\mathbb{R})$ are simply the groups of all quaternions and, respectively, pseudo-quaternions.

For the purposes of our overview and outlook, we will illustrate these general remarks by considering two plane waves
and using them to Fourier transform the momentum space constraints of Section~\ref{deformed wave equations}.

The f\/irst is def\/ined on $L_3^{+\uparrow}\simeq {\rm SL}(2,\mathbb{R})/\mathbb{Z}_2$ and is the analogue of the `bosonic' plane
wave def\/ined on ${\rm SO}(3)$~\cite{FL2,FM}.
It is the map
\begin{gather}\label{Bwave}
\psi^B_\star: \ {\rm SL}(2,\mathbb{R})/\mathbb{Z}_2
\times
\mathbb{R}^3 \rightarrow \mathbb{C},
\qquad
\psi^B_\star(u,x)=\exp(i  \epsilon(\mathcal{P}_3)  \mathcal{P}^a x_a),
\end{gather}
where $u\in {\rm SL}(2,\mathbb{R})$ is again parametrised as in~\eqref{def cP}, $\epsilon(\mathcal{P}_3)$ is the sign of
$\mathcal{P}_3$, and $x\in\mathfrak{sl}(2,\mathbb{R})^*\simeq \mathbb{R}^3$ as in Section~\ref{sec: relativistic wave equations}.
The inclusion of the sign of $\mathcal{P}_3$ means that the argument of the exponential is invariant under the
$\mathbb{Z}_2$ quotient $u\mapsto -u$ and therefore a~function on the quotient ${\rm SL}(2,\mathbb{R})/\mathbb{Z}_2$.
As in the Euclidean case, it is not def\/ined on the set of measure zero where $\mathcal{P}_3=0$.

The multiplication via the $\star$-product~\eqref{starprod} implies
\begin{gather*}
\exp\big(i   \epsilon\big(\mathcal{P}_3^{(1)}\big)\mathcal{P}_a^{(1)} x^a\big)
\star\exp\big(i \epsilon\big(\mathcal{P}_3^{(2)}\big)\mathcal{P}_a^{(2)}x^a\big)
=\exp\big(i  \epsilon\big(\mathcal{P}_3^{(1\oplus 2)}\big)\mathcal{P}_a^{(1\oplus 2)} x^a\big),
\end{gather*}
with $\mathcal{P}_3^{(1\oplus 2)}$ and $ \mathcal{P}_a^{(1\oplus 2)}$, $a=0,1,2$,
def\/ined via
\begin{gather*}
u^{(1)}u^{(2)}= \mathcal{P}_3^{(1\oplus 2)} \id + \lambda \mathcal{P}_a^{(1\oplus 2)}t_a.
\end{gather*}

Turning now to the covariant momentum space constraints def\/ining irreducible representations of
$\mathcal{D}({\rm SL}(2,\mathbb{R}))$, we note that momentum space constraints for the bosonic f\/ields~\eqref{noncommutative
Klein Gordon equation momentum space} and~\eqref{noncommutative s=1 equation momentum space} make sense for functions
$\tilde\phi$ on ${\rm SL}(2,\mathbb{R})$ which are invariant under $u\rightarrow -u$ and thus def\/ined on
${\rm SL}(2,\mathbb{R})/\mathbb{Z}_2$, while the spin $\frac 1 2 $ constraint~\eqref{noncommutative Dirac momentum space} does
not descend to the quotient.
The group Fourier transform
\begin{gather*}
\phi^B_\star (x) = \int_{{\rm SL}(2,\mathbb{R})/\mathbb{Z}_2} du \, \psi^B_\star(u,x)  \tilde\phi(u)
\end{gather*}
turns the momentum space constraints for the bosonic f\/ields~\eqref{noncommutative Klein Gordon equation momentum space}
and~\eqref{noncommutative s=1 equation momentum space} into formally the same equations as~\eqref{noncommutative Klein
Gordon equation position space}  and~\eqref{noncommutative s=1 equation position space} for $\phi^B_\star (x)$, but with
$\partial_a$ now denoting the {\em usual} partial deriva\-ti\-ve~$\partial/ \partial x^a$.

The second form of plane wave we want to consider here uses the exponential map to parametrise ${\rm SL}(2,\mathbb{R})$.
Although the exponential map $\mathfrak{sl}(2,\mathbb{R})\rightarrow {\rm SL}(2,\mathbb{R})$ is not surjective and therefore
cannot be used to parametrise the entire group, its image includes the `inside of the lightcone'
\begin{gather*}
{\rm SL}(2,\mathbb{R})_\pm=\bigcup_{\lambda m \in (-2\pi,0)\cup (0,2\pi)} C^{T}(\lambda m),
\end{gather*}
on which the elements of massive irreducible representations have their support.
For the purpose of Fourier transforming the momentum space constraints of Section~\ref{deformed wave equations}, it is
therefore suf\/f\/icient to consider elements in ${\rm SL}(2,\mathbb{R})_\pm$ for which we can def\/ine the `exponential' plane wave
\begin{gather}\label{Ewave}
\psi^E_{\star}(u,x)=\exp\big(i x^ap_a\big),
\qquad
u=\exp\big(\lambda p_at^a\big) \in {\rm SL}(2,\mathbb{R})_\pm.
\end{gather}
The associated group Fourier transform maps this to
\begin{gather*}
\phi^E_{\star}(x) =\int_{{\rm SL}(2,\mathbb{R})_\pm} du \, \psi^E_{\star}(u,x) \tilde\phi (u ).
\end{gather*}
The momentum constraints~\eqref{noncommutative Klein Gordon equation momentum space},~\eqref{noncommutative Dirac
momentum space} and~\eqref{noncommutative s=1 equation momentum space} on $\tilde\phi$ imply equations for
$\phi^E_{\star}(x)$ involving {\em exponentiated} dif\/ferential operators.

For spin $\frac12$, the left hand side of~\eqref{group constraint spin 1/2} produces the exponentiated Dirac operator
\begin{gather*}
e^{-\frac \lambda 2 \gamma^a\partial_a}\phi^E_{\star}(x),
\end{gather*}
which was considered in a~very dif\/ferent context by Atiyah and Moore in~\cite{AM}.
The authors considered dif\/ference-dif\/ferential versions of several fundamental equations of physics, including the Dirac
equation, allowing for advanced and retarded as well as advanced-retarded versions.
For spin $\frac 1 2 $, this involves in an essential way the exponential of the Dirac operator.
Their work stresses the relation between exponentiated dif\/ferential operators and dif\/ference equations, and explores the
consequences of using such equations in fundamental physics.

The appearance of dif\/ference-dif\/ferential equations in theories with curved momentum space was also pointed out
in~\cite{MW} in the context of (2+1)-dimensional gravity.
However, with few exceptions~\cite{SS3}, this point of view has not received much attention in the context of
generalised Fourier transforms and quantum groups.

\section{Conclusion and outlook}\label{Section5}

In this paper, we studied the consequences of curved momentum space for the spacetime physics of massive and spinning
particles in a~particular model.
We have kept an open mind about the motivation for studying curved momentum space.
The model considered here comes from (2+1)-dimensional quantum gravity, and has the added benef\/it of maintaing (a
deformed version of) Poincar\'e symmetry.
However, as reviewed in the Introduction, curved momentum space arises in several contexts and has been of theoretical
interest at least since Born's pioneering considerations in~\cite{Born}.

Our interest in the spacetime description of the particles via relativistic wave equations stems from the fundamental
role such equations play in relativistic quantum mechanics and quantum f\/ield theory.
The wave equations, not the equivalent momentum constraints, provide the standard route for constructing interacting
theories, be it via coupling to an external classical f\/ield (as in the Dirac equation for the electron in the hydrogen
atom) or in a~fully interacting quantum f\/ield theory.
It is therefore interesting to see how relativistic wave equations are modif\/ied when momentum space becomes curved.

In Section~\ref{noncommutative Fourier transformation}, we gave an overview over dif\/ferent ways of obtaining the wave
equations from the covariant momentum constraints derived earlier via Fourier transforms.
Our discussion there should be viewed as a~f\/irst step, pulling together relevant approaches in the literature and
preparing the ground for a~mathematically precise treatment.
At this stage, our f\/indings can be summarised as follows.

The quantum group Fourier transform is perhaps the most elegant method of Fourier transforming the covariant momentum
space constraints, and directly takes us into the realm of non-commutative geometry.
However, this route is strewn with considerable mathematical challenges.
At present, the formula~\eqref{Fourier} leads to equations like~\eqref{noncommutative Dirac position space} which are
re-interpretations of mathematically well-def\/ined momentum-space constraints but do not have an independent
mathematically rigorous def\/inition.

The group Fourier transform, by contrast, leads to equations on ordinary $\mathbb{R}^3$ which have an independent
mathematical meaning as dif\/ferential or dif\/ference-dif\/ferential equations.
However, here the nature of the equation depends on the choice of plane wave.
We have only considered two choices of plane waves here, and found conventional wave equations from the `bosonic' plane
wave~\eqref{Bwave} but dif\/ference-dif\/ferential equations from the `exponential' plane wave~\eqref{Ewave}.
The latter are suggestive of a~discrete spacetime geometry as considered in~\cite{MW} and are interesting from the more
general viewpoint emphasised in~\cite{AM}.

There are several avenues for developing the work started here.
One would like to extend the set of wave equations to include equations for massless particles and for particles with
spins other than 0, $\frac12$ and 1.
Anyonic excitations are relevant and interesting in this context.
They arise naturally in the context of 2+1 gravity, where the spin is quantised in units which depend on the
mass~\cite{BMS}.
To study them one needs to work with the universal cover of ${\rm SL}(2,\mathbb{R})$.

It would also be interesting to consider both the quaternionic extension $\mathbb{R}^+
\times
{\rm SL}(2,\mathbb{R})$ and a~Lorentzian version of the spinorial description in~\cite{DGL} in future work.
This will presumably lead to yet dif\/ferent types of `wave equations'.

On general grounds, the choice of coordinates on the group ${\rm SL}(2,\mathbb{R})$ should not matter, and one expects plane
waves which dif\/fer only by dif\/ferent coordinates to lead to equivalent wave equations after Fourier transform.
However, in so far as dif\/ferent plane waves ref\/lect a~dif\/ferent choice of group (e.g.~${\rm SL}(2,\mathbb{R})/\mathbb{Z}_2$ versus ${\rm SL}(2,\mathbb{R})$ or $\mathbb{R}^+
\times
{\rm SL}(2,\mathbb{R})$), one would expect the associated wave equations to be dif\/ferent.

Ultimately, one would like to achieve a~systematic understanding of the nature and inter-relationship of wave equations
that can be obtained via Fourier transform of a~given curved momentum space constraint.

\appendix

\section[Basis and f\/inite-dimensional representations of $\mathfrak{sl}(2,\mathbb{R})$]{Basis and f\/inite-dimensional representations of $\boldsymbol{\mathfrak{sl}(2,\mathbb{R})}$}\label{conventions}

In the main text, we use the the basis $\{t^a,\, a=0,1,2\}$ of $\mathfrak{sl}(2,\mathbb{R})$ with
\begin{gather}\label{fundrep}
t^0=\frac{1}{2}
\begin{pmatrix}
\phantom{-} 0 & 1
\\
-1 & 0
\end{pmatrix}
,
\qquad
t^1=\frac{1}{2}
\begin{pmatrix}
1 & \phantom{-}0
\\
0 & -1
\end{pmatrix}
,
\qquad
t^2=\frac{1}{2}
\begin{pmatrix}
0 & 1
\\
1 & 0
\end{pmatrix}
.
\end{gather}
The basis elements satisfy
\begin{gather}\label{ta tb}
t^at^b= -\frac{1}{4}\eta^{ab}\id+\frac{1}{2}\epsilon^{abc}t_c,
\end{gather}
where $\id$ denotes the 2 $
\times
$ 2 identity matrix.
As a~result, we have the commutation relations
\begin{gather}\label{commutation relations t's}
\big[t^a,t^b\big]= \epsilon^{abc}t_c
\end{gather}
and the anticommutation relations
\begin{gather}\label{anticommutation relations t's}
\big\{t^a,t^b\big\}=t^at^b+t^bt^a= -\frac{1}{2}\eta^{ab}\id.
\end{gather}
Finally, we note the orthogonality relations
\begin{gather*}
-2\tr\big(t^at^b\big)=\eta^{ab}.
\end{gather*}

The representation theory of $\mathfrak{sl}(2,\mathbb{R})$ is best studied in terms of raising and lowering operators
\begin{gather}\label{def HE+E-}
H=t^1,
\qquad
E_+=t^2+t^0=
\begin{pmatrix}
0 & 1
\\
0 & 0
\end{pmatrix}
,
\qquad
E_-=t^2-t^0 =
\begin{pmatrix}
0 & 0
\\
1 & 0
\end{pmatrix}
,
\end{gather}
with commutation relations
\begin{gather*}
[H,E_+]= E_+,
\qquad
[H,E_-]= -E_-,
\qquad
[E_+,E_-]= 2H.
\end{gather*}
It is well-known~\cite{Sternberg} that the f\/inite-dimensional representations of the Lie algebra
$\mathfrak{sl}(2,\mathbb{R})$ are parametrised by $j \in \frac{1}{2}(\mathbb{N} \cup {0})$.
For each value of~$j$ there is a~unique irreducible representation~$\rho^j$ on~$V_j\simeq \mathbb{C}^{2j+1}$.
The standard basis $\{w_j, w_{j-1}, \dots, w_{1-j}, w_{-j}\}$ of $V_j$ is such that
\begin{gather*}
\rho^j(H) w_k = k w_{k},
\qquad
\rho^j(E_-) w_k = (j+k) w_{k-1},
\qquad
\rho^j(E_+) w_k = (j-k) w_{k+1}.
\end{gather*}
These representations are not unitary.
Only $\rho^j(t^0)$ has imaginary eigenvalues and exponentiates to a~unitary matrix.
In the main text, we work with an eigenbasis
\begin{gather}\label{eigenbasis}
\{|j,j\rangle,|j,j-1\rangle,\ldots, |j,1-j\rangle, |j,-j\rangle\}
\end{gather}
of $\rho^j(t^0)$ satisfying $\rho^j(t^0)\lvert j,k \rangle = i k | j,k\rangle$.

\subsection*{Acknowledgements}

MW thanks the Department of Mathematics at Heriot-Watt University for hospitality during a~six-months visit in 2010 when
the bulk of the research reported here was carried out.
BJS thanks Sergio Inglima for discussions and comments on a~draft version of the manuscript.
Both MW and BJS thank the University of Ghana for hospitality during a~research visit in April 2010.

\pdfbookmark[1]{References}{ref}
\LastPageEnding

\end{document}